\documentclass[preprint,prb,showpacs]{revtex4}
\usepackage{graphicx}

\begin{document}

\title{Stable Magnetostatic Solitons in Yttrium Iron
Garnet Film Waveguides for Tilted in-Plane Magnetic Fields}

\author{R.\ Khomeriki} 
\thanks{khomeriki@hotmail.com}

\affiliation{
Department of Physics and Astronomy, The University of 
Oklahoma, OK 73019, USA; \\ 
Department of Physics, Tbilisi
State University, Chavchavadze ave.\ 3, Tbilisi 380028, GEORGIA}

\author{L.\ Tkeshelashvili}
\thanks{lasha@tkm.physik.uni-karlsruhe.de}
 
\affiliation{Institut f\"ur Theorie der 
Kondensierten Materie,
University of Karlsruhe,
P.O. Box 6980, 76128 Karlsruhe, GERMANY;
\\ Institute of
Physics, Tamarashvili str.\ 6, Tbilisi 380077, GEORGIA}

\begin{abstract}

The possibility of nonlinear pulses generation in
Yttrium Iron Garnet thin films for arbitrary direction between waveguide
and applied static in-plane magnetic field is considered. Up to now only the
cases of in-plane
magnetic fields either perpendicular or parallel to waveguide direction
have been studied both experimentally and theoretically. In the present
paper it is
shown that also for other angles (besides 0 or 90 degrees) between
a waveguide and static in-plane magnetic field the stable bright
or dark (depending on magnitude of magnetic field) solitons could be 
created.
\end{abstract}

\vspace{5mm}


\maketitle

\section {Introduction}

The investigation of magnetostatic envelope solitons in yttrium-iron garnet 
thin
magnetic films is one of the "hot topics" in nowadays physics.  The advanced
instrumentation for microwave pulse generation, detection and analysis
together with the solid theoretical base
 has led to a growing interest in 
studying such
localized objects.
The definition "magnetostatic soliton" refers to a propagating pulse formed 
by large
wavelength spin excitations which do not "feel" the exchange interaction and
only the dipolar interactions could be taken into account. Thus the 
processes are
characterized by the Landau-Lifshitz and magnetostatic equations.

The linearized solutions of these equations
were obtained by Damon and Eshbach \cite{damon} 40 years
ago for arbitrary direction between wave vector of spin excitations and 
in-plane
magnetic field. The nature of those excitations has been also studied 
experimentally
\cite{hurben}. The weakly nonlinear limit for the mentioned equations also 
was considered
for the particular cases when wave vector of spin excitations is either 
parallel
(backward volume waves) or perpendicular (surface waves) to the direction of 
in-plane
magnetic field. It was found out \cite{zvezdin} that the envelope of spin 
excitations
in both cases satisfy 2D Nonlinear Schr\"odinger (NLS) equation which 
permits well-known
1D soliton solutions depending on the relative sign of dispersion and 
nonlinear terms.

In full accordance with the theoretical predictions the bright solitons have 
been observed
for nonlinear backward volume waves case \cite{boyle} - \cite{xia} (in-plane 
field is directed
parallel to the carrier wave vector and propagation velocity of envelope 
soliton), while the
dark solitons are created in case of nolinear surface waves 
\cite{chen2}-\cite{kalinikos}
(carrier wave vector and group velocity is perpendicular to the magnetic 
field). It should
be especially noted that all the mentioned solitons are observed in narrow 
strips. In such
geometries the transverse instabilities do not develop and experiments show 
the stable
propagation of 1D solitons along the waveguides. At the same time, in wide 
samples 1D solitons
are in general unstable \cite{zakharov}, \cite{kivshar} and form metastable 
spin wave bullets
which  decay either after edge reflection or mutual interaction 
\cite{zalit1}, \cite{zalit2}.

We emphasize that the solitons in in-plane magnetized films are studied both 
theoretically and
experimentally only for two particular cases when the pulse propagates along 
or perpendicular
to the magnetic field. Only very recently the general case of linear and
nonlinear magnetostatic wave propagation in wide samples was investigated
\cite{ref}
for a wide range of angles between propagation velocity and magnetic field.
In this connection the natural
question arises: why one does not consider the nonlinear pulses in 
waveguides which are not
either parallel or perpendicular to in-plane
static magnetic field. As we show below for each magnitude of internal 
magnetic field it is
possible to choose the direction (besides 0 or 90 degrees) of waveguide 
respect
to magnetic field direction for which
stable propagation of envelope solitons is allowed (see inset of Fig. 1 for 
a geometry of
the problem). We determine the limits for magnitude of magnetic field, angle 
between
waveguide and magnetic field vector and pulse frequency necessary for 
creation of dark or
bright envelope solitons. We also calculate their widths and propagation 
velocities and
claim that such solitons could be experimentally observed.

\section{Basic Consideration: Linear Magnetostatic Waves}

The linear consideration is based upon the Damon-Eshbach formulation
\cite{damon} and its generalization by Hurben and Patton \cite{hurben}
for the case of arbitrary angles between a wave vector $\vec{k}$ and
static internal magnetic field $\vec{H}_0$.
Further we will examine only so-called "near-uniform" case ($kd\ll 1$,
$d$ stands for a film thickness) and derive the dispersion expansion
over the parameter $kd$ up to a second order. Therefore we present here
only the steps necessary for this purpose. Consideration of the mentioned
wave number range sufficiently simplifies calculations and, besides that,
most of the experiments on the magnetostatic envelope solitons are made 
having
such carrier wave numbers.

Examining an in-plane magnetized ferromagnetic film with unpinned
surface spins let us make the following definitions: $z$ is a direction of 
internal static
magnetic field; $\vec r$ indicates the radius
vector lying in the sample plane ($y,~z$) and $x$ is a coordinate along
the direction perpendicular to the film plane.
Then one can write down the Landau-Lifshitz and magnetostatic equations:
\begin{equation}
\frac{d{\vec M}}{dt}=-g[{\vec M}\times{\vec H}] \qquad
div({\vec H}+4\pi{\vec M})=0 \qquad rot{\vec H}=0.  \label{2}
\end{equation}
Here $g$ is a modulus of the gyromagnetic ratio for electrons; $\vec{M}$ is
a magnetization
density vector and $\vec{H}$ is an internal total magnetic field. 
Introducing
dynamical dimensionless quantities $\vec{m}=(\vec{M}-\vec{M}_0)/M_0$
and $\vec{h}=(\vec{H}-\vec{H}_0)/H_0$ the following equations are obtained 
in
the linear limit over $|\vec{m}|$:
$$
\frac{dm_x}{dt}=\omega_H(m_y-h_y); \qquad
\frac{dm_y}{dt}=-\omega_H(m_x-h_x);
$$
\begin{equation}
\frac{\partial}{\partial x}\left(\omega_Hh_x+\omega_Mm_x\right)+
\frac{\partial}{\partial y}\left(\omega_Hh_y+\omega_Mm_y\right)+
\frac{\partial}{\partial z}\left(\omega_Hh_z\right)=0, \label{3}
\end{equation}
where $\omega_H=gH_0$ and $\omega_M=4\pi gM_0$. Defining $\vec{h}=\vec{grad}
\Phi$ and searching for the solution of Eq. (\ref{3}) in periodical form 
over $t$ (i.e. $\Phi$ and $m_{x,y}$ being proportional to 
$exp(-i\omega t)$) we get:
\[
\frac{\partial^2}{\partial z^2}\Phi 
+(\chi_1+1)\left(\frac{\partial^2}{\partial
y^2}+\frac{\partial^2}{\partial x^2}\right)\Phi=0 \qquad \mbox{for} \qquad
|x|<\frac{d}{2},
\]
\begin{equation}
\left(\frac{\partial^2}{\partial z^2}+\frac{\partial^2}{\partial y^2}+
\frac{\partial^2}{\partial x^2}\right)\Phi=0 \qquad \mbox{for} \qquad
|x|>\frac{d}{2} \label{4}
\end{equation}
and
\[
m_x=\frac{\omega_H}{\omega_M}\left(i\chi_2\frac{\partial\Phi}{\partial y}+
\chi_1\frac{\partial\Phi}{\partial x}\right), \qquad
m_y=-\frac{\omega_H}{\omega_M}\left(i\chi_2\frac{\partial\Phi}{\partial x}-
\chi_1\frac{\partial\Phi}{\partial y}\right),
\]
where
\[
\chi_1=\frac{\omega_H\omega_M}{\omega_H^2-\omega^2}, \qquad
\chi_2=\frac{\omega\omega_M}{\omega_H^2-\omega^2}.
\]
Thus from (\ref{4}) we can write down a linear solution of (\ref{2}) in the
form \cite{hurben}:
\[
\Phi=\left(Ae^{\kappa x}+Be^{-\kappa x}\right)e^{-i(\omega 
t-\vec{k}\vec{r})}
\qquad \mbox{for} \qquad |x|<\frac{d}{2};
\]
\begin{equation}
\Phi=Ce^{-k(x-d/2)}e^{-i(\omega t-\vec{k}\vec{r})}
\qquad \mbox{for} \qquad x>\frac{d}{2}; \label{5}
\end{equation}
\[
\Phi=De^{k(x+d/2)}e^{-i(\omega t-\vec{k}\vec{r})}
\qquad \mbox{for} \qquad x<-\frac{d}{2},
\]
where $A$, $B$, $C$ $D$ are arbitrary constants at the present stage,
\[
\kappa^2=\frac{k^2+\chi_1 k^2_y}{1+\chi_1}
\]
and let us remind that two dimensional vectors $\vec{r}=(y,~z)$ and
$\vec{k}=(k_y,~k_z)$ lye in the film plane and $k\equiv\sqrt{k_y^2+k_z^2}$.

If $\kappa^2>0$ we deal with a so-called surface mode, otherwise $\kappa=
i\sqrt{-\kappa^2}$ and a volume mode exists. However, note that due to the
condition of "near-uniformity" $kd\ll 1$ the difference between these two
modes is negligible.

The dispersion relation can be obtained from (\ref{5}) if we remember about
the boundary conditions. Particularly the functions $\Phi$ and $h_x+4\pi 
m_x$
should be continuous on the boundaries $-d/2$ and $d/2$ of the film.
The dispersion relation for both modes could be written as follows:
\begin{equation}
2k(\chi_1+1)\kappa\frac{e^{\kappa d}+e^{-\kappa d}}{e^{\kappa d}-e^{-\kappa 
d}}
-k_y^2\chi_2^2+k^2+\kappa^2(\chi_1+1)^2=0. \label{6}
\end{equation}
Working in the limit $kd\ll 1$ and keeping only the terms up to the second 
order
of this parameter the following expression is obtained:
\begin{equation}
\omega=\omega_0+\frac{\omega_M}{4\omega_0}\frac{d}{k}\left(\omega_Mk_y^2-
\omega_Hk_z^2\right)-\frac{\omega_M^2}{32\omega_0^3}\frac{d^2}{k^2}
\left(\omega_Mk_y^2-
\omega_Hk_z^2\right)^2+\frac{\omega_M}{4\omega_0}d^2\left(\omega_H\frac{k_z^2}{3}-
\omega_Mk_y^2\right), \label{7}
\end{equation}
where $\omega_0\equiv\omega(k=0)=\sqrt{\omega_H(\omega_H+\omega_M)}$. Then 
we get
from (\ref{7}) the following expressions for the derivatives of $\omega$ 
over
$k_y$ and $k_z$:
\[
v_y=\frac{\partial
\omega}{\partial k_y}=\frac{\omega_M}{4\omega_0}d\left\{\frac{k_y}{k}\left[
\omega_M\frac{k_y^2}{k^2}+(2\omega_M+\omega_H)\frac{k_z^2}{k^2}\right]
+{\cal O}_1(kd)\right\},
\]
\[
v_z=\frac{\partial
\omega}{\partial k_z}=-\frac{\omega_M}{4\omega_0}d\left\{\frac{k_z}{k}\left[
\omega_H\frac{k_z^2}{k^2}+(2\omega_H+\omega_M)\frac{k_y^2}{k^2}\right]
+{\cal O}_2(kd)\right\},
\]
\[
\omega''_{yy}=\frac{\partial^2\omega}{\partial k_y^2}
=\frac{\omega_M}{4\omega_0}\frac{d}{k}\left\{\frac{k_z^2}{k^2}\left[
(2\omega_M+\omega_H)\frac{k_z^2}{k^2}-(2\omega_H+\omega_M)\frac{k_y^2}{k^2}
\right]+{\cal O}'_1(kd)\right\},
\]
\[
\omega''_{zz}=\frac{\partial^2\omega}{\partial k_z^2}
=\frac{\omega_M}{4\omega_0}\frac{d}{k}\left\{\frac{k_y^2}{k^2}\left[
(2\omega_M+\omega_H)\frac{k_z^2}{k^2}-(2\omega_H+\omega_M)\frac{k_y^2}{k^2}
\right]+{\cal O}'_2(kd)\right\},
\]
\begin{equation}
\omega''_{yz}=\frac{\partial^2\omega}{\partial k_zk_y}
=-\frac{\omega_M}{4\omega_0}\frac{d}{k}\left\{\frac{k_yk_z}{k^2}\left[
(2\omega_M+\omega_H)\frac{k_z^2}{k^2}-(2\omega_H+\omega_M)\frac{k_y^2}{k^2}
\right]+{\cal O}'_3(kd)\right\}. \label{8}
\end{equation}
The higher approximation terms ${\cal O}_j(kd)$ and ${\cal O}'_j(kd)$
are not presented here because of their
rather cumbrous form, but we use them in the calculation as far as the 
leading
terms in expressions (\ref{8}) vanish in vicinity of some points, e.g. 
$k_y=0$ or
$k_z=0$.

\section{Weakly Nonlinear Limit: Soliton Solutions}

Defining a wave envelope $u$
\[
m_x+im_y=u\cdot e^{-i(\omega t-\vec{k}\vec{r})}
\]
and following the well known modulation approach \cite{karpman}, 
\cite{zvezdin}
the nonlinear equation for wave envelope $u$ is derived (we 
redirect
reader for details of obtaining this equation to the recent paper Ref. 
\cite{boyle}
where the full procedure is well described):
\begin{equation}
i\left(\frac{\partial u}{\partial t}+v_y\frac{\partial u}{\partial y}+
v_z\frac{\partial u}{\partial z}\right)+\frac{\omega''_{yy}}{2}
\frac{\partial^2 u}{\partial y^2}+\frac{\omega''_{zz}}{2}
\frac{\partial^2 u}{\partial z^2}+\omega''_{yz}
\frac{\partial^2 u}{\partial y\partial z}-N|u|^2u=0. \label{10}
\end{equation}
All of the coefficients are defined by formulas (\ref{8}) except the 
nonlinear coefficient
$N$ which could be easily calculated taking into account that in nonlinear 
case we
have a following identity:
\[
\omega_M=4\pi gM_0m_z.
\]
Then Substituting in (\ref{7}) the expansion of $m_z$ in weakly nonlinear 
limit
$m_z=1-|u|^2/2$ and using
expression for $N$ from Refs. \cite{karpman}, \cite{zvezdin}, \cite{boyle}
we get in large wavelength limit ($kd\ll 1$)
\begin{equation}
N=\frac{\partial\omega}{\partial |u|^2}\biggr|_{k\rightarrow 
0,~|u|\rightarrow 0}=
\frac{\partial\omega_0}{\partial |u|^2}\biggr|_{|u|\rightarrow 0}
=-\frac{\omega_H\omega_M}{4\omega_0}. \label{9}
\end{equation}
Let us mention that if carrier wave vector is parallel or perpendicular to 
the
static internal magnetic field the coefficients $v_y$ and $w''_{yz}$ are 
equal to zero
\cite{zvezdin}, \cite{boyle}.
But for arbitrary angles between $\vec{k}$ and $\vec{H_0}$ that is not the 
case.
Therefore we should introduce a new frame of references in order to 
vanish
the nondiagonal term with coefficient $\omega''_{yz}$.
This could be done rotating the frame of references $yz$ by the angle 
$\vartheta$
\begin{equation}
\xi=z\cos\vartheta+y\sin\vartheta, \qquad 
\eta=y\cos\vartheta-z\sin\vartheta, \label{11a}
\end{equation}
where
\begin{equation}
tg2\vartheta=2\frac{\omega''_{yz}}{\omega''_{zz}-\omega''_{yy}}. \label{11}
\end{equation}
Then from (\ref{10}) - (\ref{11}) we obtain the following nonlinear 
equation:
\begin{equation}
i\left(\frac{\partial u}{\partial t}+v_1\frac{\partial u}{\partial \xi}+
v_2\frac{\partial u}{\partial \eta}\right)+\frac{1}{2}R
\frac{\partial^2 u}{\partial \xi^2}+\frac{1}{2}S
\frac{\partial^2 u}{\partial \eta^2}-N|u|^2u=0, \label{12}
\end{equation}
where
\[
R=\omega''_{zz}cos^2\vartheta+2\omega''_{zy}cos\vartheta\sin\vartheta
+\omega''_{yy}sin^2\vartheta,
\]
\begin{equation}
S=\omega''_{zz}sin^2\vartheta-2\omega''_{zy}cos\vartheta\sin\vartheta
+\omega''_{yy}cos^2\vartheta, \label{R}
\end{equation}
\begin{equation}
v_1=v_zcos\vartheta+v_ysin\vartheta, \qquad 
v_2=v_zsin\vartheta-v_ycos\vartheta. \label{v}
\end{equation}
Afterwards in the moving frame of references
\[
\xi_1=\xi -v_1t \qquad \eta_1= \eta-v_2t
\]
we come to the 2D nonlinear Schr\"odinger equation:
\begin{equation}
i\frac{\partial u}{\partial t}+\frac{1}{2}R
\frac{\partial^2 u}{\partial \xi_1^2}+\frac{1}{2}S
\frac{\partial^2 u}{\partial \eta_1^2}-N|u|^2u=0 \label{12a}
\end{equation}
and can write down its 1D bright or dark soliton solutions assuming that 
soliton
envelope is a function only of variables $\xi_1$ and $t$ (thus soliton 
propagates along a spatial axis $\xi$ with a velocity $v_1$). If $NR<0$ we 
have bright soliton
with envelope
\begin{equation}
|u|=|u|_{max}sech\left\{\frac{\xi_1}{\Lambda}\right\}, \qquad
\label{13}
\end{equation}
while in case $NR>0$ dark soliton solution is permitted:
\begin{equation}
|u|=|u|_{max}\left|\frac{\sqrt{1-A^2}}{A}+i\cdot 
tanh\left\{\frac{\xi_1}{\Lambda}\right\}\right|,
\label{13a}
\end{equation}
where $A$ denotes the contrast of dark soliton (if $A=1$ one has a black 
dark soliton and gray
dark otherwise) and soliton width $\Lambda$ is defined for both cases by the 
same way:
\begin{equation}
\Lambda=\left|\frac{R}{N}\right|^{1/2}\frac{1}{|u|_{max}}, \label{13b}
\end{equation}
Now we shall discuss the question about the stability of these 1D solitons.

\section{Stable Solitons in Waveguides for Tilted Magnetic Fields}

As well known 1D soliton solutions (\ref{13}) and (\ref{13a}) of 2D NLS are 
not stable to
the transverse modulations with wavenumbers $0<\kappa<\kappa_c$. According 
to the recent
results (see e.g. Ref. \cite{kivshar}) $\kappa_c\sim 1/\Lambda$, thus, if 
the limits of
transverse variable $\eta_1$
is less than soliton width the instabilities do not develop and 1D soliton 
solutions
(\ref{13}) and (\ref{13a}) would be stable. When one has fully spatial 
transverse variable
$\eta_1$ the
above condition means that narrow samples should be used. In our case we 
have the mixed
variable $\eta_1=\eta-v_2t$ and therefore also the condition for time 
dependent part has
to be introduced: $v_2t<\Lambda$ and in case
\begin{equation}
v_2=0 \label{v2}
\end{equation}
the transverse instabilities do not develop even for infinite time. Thus 
besides the condition
(\ref{11}) we get from (\ref{v}) and (\ref{v2}) additional condition on the 
stable soliton
parameters:
\begin{equation}
tg\vartheta=\frac{v_y}{v_z}. \label{cond2}
\end{equation}
Afterwards, in view of both conditions (\ref{v}) and (\ref{cond2}) we 
finally obtain the following
equality:
\begin{equation}
\frac{\omega''_{yz}}{\omega''_{zz}-\omega''_{yy}}=\frac{v_yv_z}{v_z^2-v_y^2}. 
\label{cond0}
\end{equation}

Solving (\ref{cond0}) as an expansion over small parameter $kd$ we simply 
come to the following
expression for the angle $\varphi$ between carrier wavevector ${\vec k}$ and static 
magnetic field:
\begin{equation}
\sin\varphi=-\sqrt{\frac{\omega_H}{\omega_H+\omega_M}}\left(1+\frac{\omega_M
(3\omega_H+\omega_M)}{3(\omega_H^2-\omega_M^2)}kd\right). \label{80}
\end{equation}
Obviously there exist also trivial solutions $\varphi=0$ and $\varphi=90^o$ 
which will
not be considered as long as they correspond to the well studied cases of 
bright
backward volume
wave and dark surface wave solitons, respectively. Further using the 
condition (\ref{cond2})
and definitions for dispersion coefficient (\ref{R}) we get the expressions 
for $\vartheta$ and $R$ as functions of expansion parameter $kd$:
\begin{equation}
\sin\vartheta=\sqrt\frac{\omega_M}{\omega_M+\omega_H}\left(1-\frac{\omega_H}{\omega_H+\omega_M}kd\right); 
\qquad
R=\frac{\omega_M(\omega_M-\omega_H)}{2\sqrt{\omega_H(\omega_M+\omega_H)}}\frac{d}{k}. 
\label{88}
\end{equation}
As long as the nonlinear coefficient $N$ according to (\ref{9}) is always 
negative the possibility of appearance of dark or bright solitons depends on 
the sign of dispersion coefficient $R$. In view of the second relation in 
(\ref{88}) we can conclude that bright solitons appears if 
$\omega_H<\omega_M$
while dark solitons could be created for larger magnetic fields 
$\omega_H>\omega_M$. From Exps.
(\ref{13b}) and (\ref{7}) we are also able
to get the expressions for soliton width and detuning of pulse frequency, 
respectively:
\begin{equation}
\Lambda=\frac{d}{|u|_{max}}\sqrt{\frac{2|\omega_M-\omega_H|}{\omega_Hkd}}; 
\qquad
\omega-\omega_0=\frac{\omega_M}{3}\sqrt{\frac{\omega_M}{\omega_H+\omega_M}}\frac{\omega_M}
{\omega_H-\omega_M}(kd)^2, \label{98}
\end{equation}
while the soliton propagation velocity could be given by simple approximate 
formula:
\begin{equation}
v\equiv v_1\simeq \frac{\omega_M 
d}{2}\sqrt{\frac{\omega_M}{\omega_M+\omega_H}}
\end{equation}

It should be noted that our perturbative approach violates if 
$\omega_H\rightarrow\omega_M$.
Besides that, we have a following restriction on the internal
static 
magnetic field: $\omega_H>0.3\omega_M$.
Otherwise the threshold for three magnon processes is reached and localized 
nonlinear wave will decay rapidly \cite{zvezdin}.

As we see all of the quantities $\vartheta$, $\omega-\omega_0$ and $\Lambda$
specifying the soliton are the functions of $kd$ and $h=\omega_H/\omega_M$.
Thus it is possible to plot $\omega-\omega_0$ and $\Lambda$ as the functions 
of $\vartheta$ for different $h$ (see Figs. 1 and 2).

In Fig. 1 we present how to choose the sample geometry (in other words how 
to choose an
angle $\vartheta$ between waveguide and static magnetic field) and frequency 
of applied pulse
for various $h\equiv\omega_H/\omega_M$ in order to create bright or dark 
soliton. While in
Fig. 2 we show the dependence of soliton width on the geometry of the 
problem and static
magnetic field. In both cases the curves are limited because of restriction 
of
"near uniformity" $kd\ll 1$ and the following parameters for YIG film are 
used: $d=10\mu m$
and $\omega_M=1750$ Oe. Note that bright solitons appear when the angle 
between waveguide and static magnetic field is less than 45$^o$. Besides 
that, the detuning of pulse should be positive in order to create dark 
solitons. For the purpose to create bright solitons the angles should be 
larger than $45^o$ and detuning has to be negative.

\section{Conclusions and Possible Experimental Setup}

Summarizing we can declare that the possibility of stable magnetostatic 
soliton propagation in in-plane magnetized ferromagnetic films in presence 
of tilted
(from waveguide direction) static in-plane magnetic fields is proved. The 
widths and velocities as well as the range of 
angles between waveguide and magnetic field
is obtained for which the stable soliton propagation is 
allowed.

However, As it was mentioned by the anonymous referee the waveguide border 
(see dashed line in the inset of Fig. 1) could 
cause the reflection of carrier wave (wave vector $\vec k$) what will change the
group velocity destructing thus the soliton. To avoid such a possibility we
propose to use tube like magnetic waveguides (see Fig. 3). Then the carrier 
wave will not be reflected and, besides that, the condition of quasi-one
dimensionality still holds. Let us make the following choice of the parameters 
of the problem: tube diameter $L=0.5$mm; film thickness $d=10\mu m$ and
$1/L\ll k\ll 1/d$. Thus the near uniformity condition $kd\ll 1$ is still valid and
simultaneously the consideration of the sample as locally flat is allowed
proving thus approximate validity of solutions like Exp. (\ref{5}). 

\vspace{1cm}

{\bf Acknowledgements:} We would like to thank Dr. David Tskhakaia for the
guiding advice.  R. Kh. acknowledges NSF-NATO award No DGE-0075191 providing 
the financial
support for his stay in Oklahoma University where the paper was completed.

\newpage

\begin{figure}[htp]
\begin{center}\leavevmode
\includegraphics[width=0.9\linewidth]{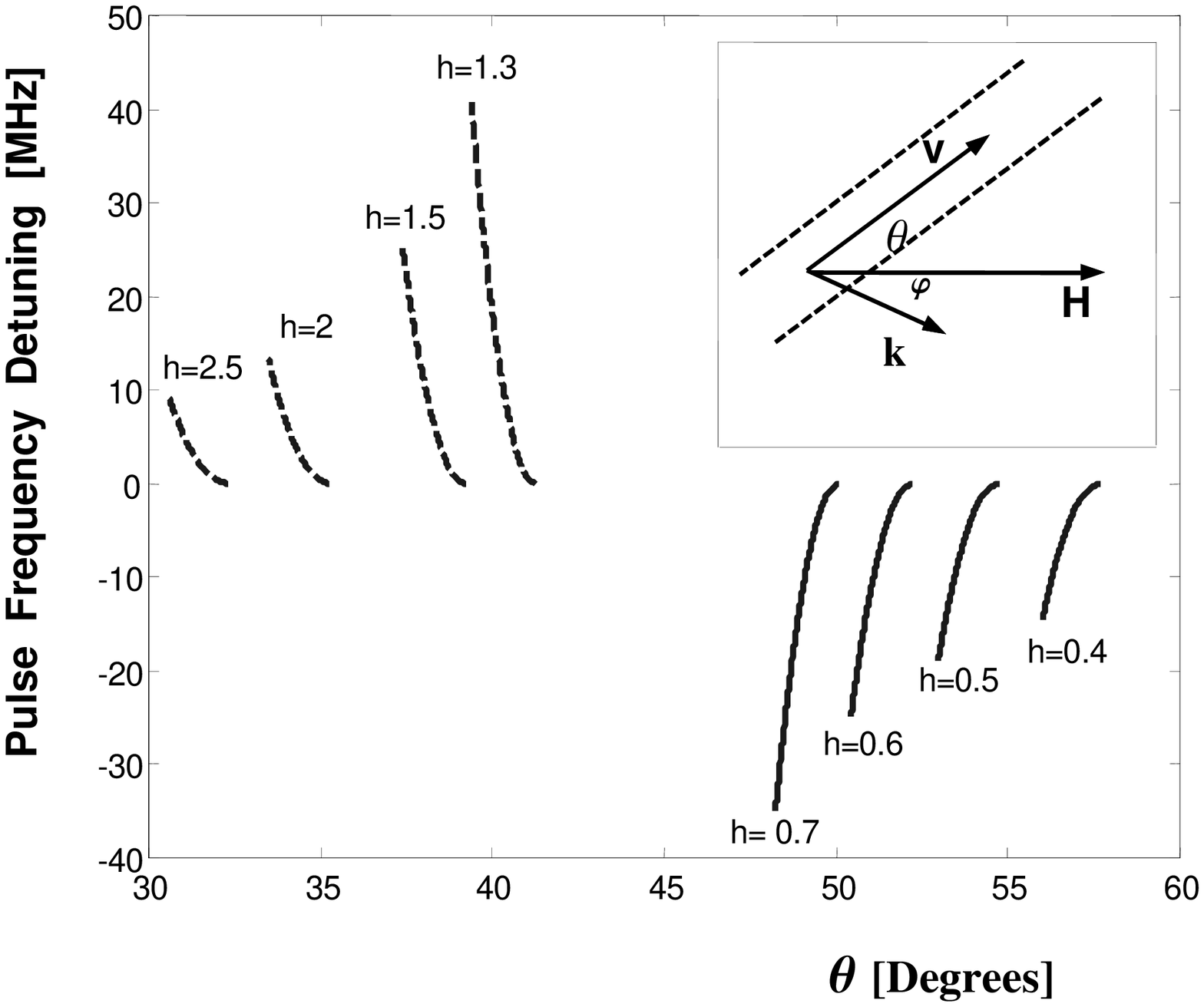}
\vspace{-10cm}
\caption{Detuning of the pulse frequency  $\omega-\omega_0$
versus an angle $\theta$ between waveguide and static in-plane magnetic field 
for its different magnitude $h=\omega_H/\omega_M$. Dashed and solid lines 
correspond to the dark and bright soliton cases, respectively. Inset shows 
the geometry of the problem and dashed lines indicate the direction of a 
waveguide. The following film parameters are used in calculations: Film 
thickness $d=10\mu m$ and value of demagnetizing field $H_M=1750$ Oe.}
\label{papa.eps}\end{center}\end{figure}

\begin{figure}[htp]
\begin{center}\leavevmode
\includegraphics[width=0.9\linewidth]{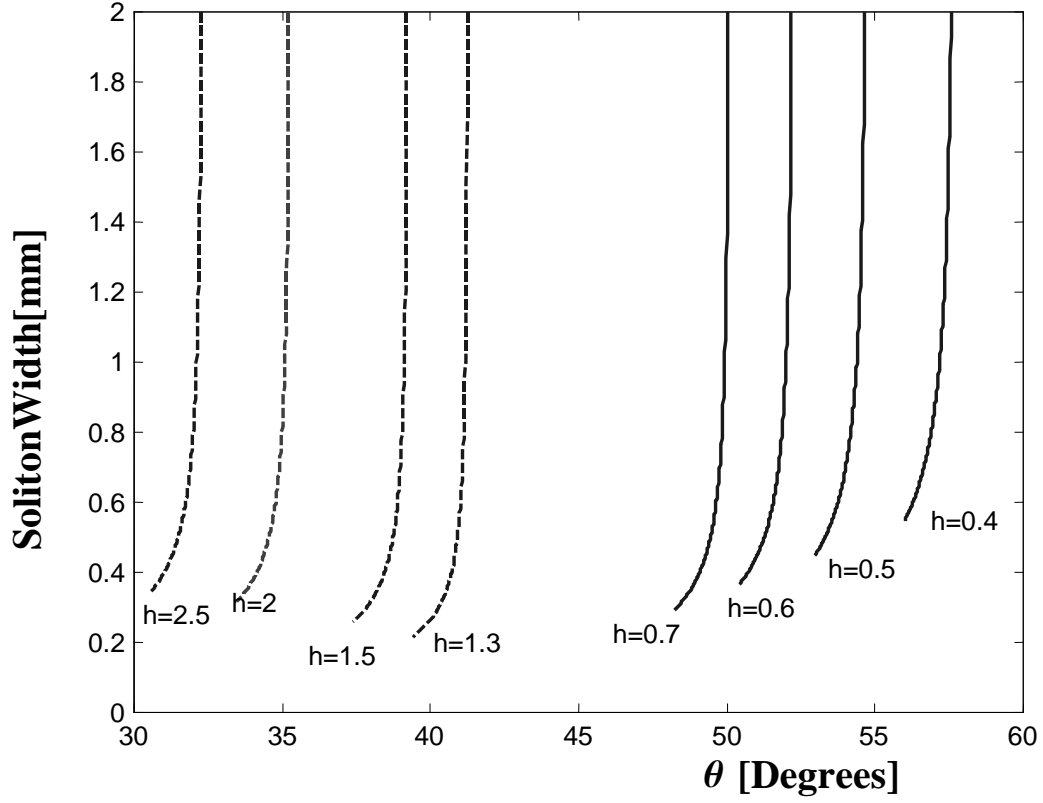}
\vspace{-5cm}
\caption{Dependence of the envelope soliton width on the angle $\theta$ between
waveguide direction and in-plane magnetic field for its different value.
Dashed and solid lines indicate dark and bright soliton cases, respectively.
As in the previous figure $h=\omega_H/\omega_M$; $H_M=1750$ Oe; film thickness
is equal to $d=10\mu m$ and, besides that, relative amplitude of soliton is
taken as follows: $|u_{max}|=0.1$.}
\label{mamt.eps}\end{center}\end{figure}

\begin{figure}[htp]
\begin{center}\leavevmode
\includegraphics[angle=-90,width=0.9\linewidth]{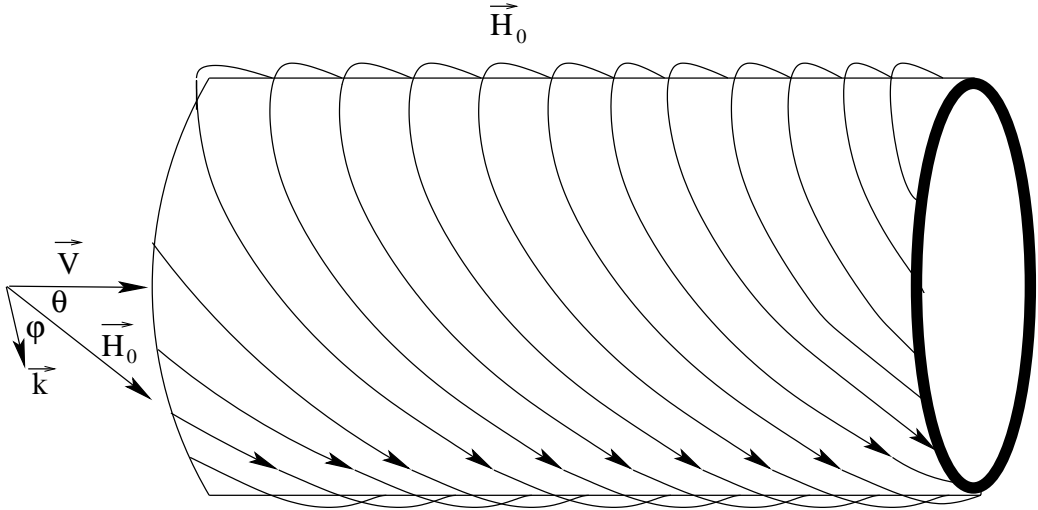}
\caption{The possible experimental setup for observation of stable magnetostatic
solitons. In each point of the sample magnetic field is parallel to the tube
surface and tilted by the angle $\theta$ from the tube symmetry axis which
is parallel to the envelope soliton propagation direction.}
\label{mamb.eps}\end{center}\end{figure}

\end{document}